\documentstyle[preprint,aps,floats]{revtex}
\input epsf.tex
\def\DESepsf(#1 width #2){\epsfxsize=#2 \epsfbox{#1}}
\def\etal{ {\em et al.}}

\def\be {\begin{equation}}
\def\ee {\end{equation}}
\def\barr{\begin{array}}
\def\earr{\end{array}}
\def\dis{\displaystyle}
\def\ra{\rightarrow}

\def\bra {\langle}
\def\ket{\rangle}

\def\l {\lambda}

\def\rpv {R_p\!\!\!\!\!\!/~~}
\def\hlp {H_{eff}^{\lambda'}}

\def\bpetapk {B^{\pm}\rightarrow \eta' K^{\pm}}
\def\betak {B\rightarrow \eta K}
\def\betak0 {B^{0}\rightarrow \eta' K^{0}}

\def\betapkstr0 {B^{0}\rightarrow \eta' K^{*0}}
\def\bpetak0 {B^{0}\rightarrow \eta' K^{0}}
\def\bpphik {B^{\pm}\rightarrow \phi K^{\pm}}

\def\msnus {m_{\tilde\nu_{iL}}^2}
\def\msells {m_{\tilde e_{iL}}^2}

\def\rtwo {\sqrt{2}}
\def\lappeq{\mathrel{\rlap{\raise.5ex\hbox{$<$}}
                    {\lower.5ex\hbox{$\sim$}}}}

\begin{document}
\preprint{\vbox{\hbox{CTP-TAMU-01-02}\hbox{}\hbox{}}}
\draft
\title{Charmless Non-Leptonic $B$ Decays and \\
R-parity Violating Supersymmetry}

\author{ B. Dutta$^1$\footnote{b-dutta@rainbow.physics.tamu.edu}, ~~
C.~S. Kim$^2$\footnote{cskim@mail.yonsei.ac.kr,~~~
http://phya.yonsei.ac.kr/\~{}cskim/}~~ and~~ Sechul
Oh$^2$\footnote{scoh@phya.yonsei.ac.kr}  }

\address{
$^1$Center For Theoretical Physics, Department of Physics, Texas A$\&$M
University,\\ College Station, TX 77843-4242\\
$^2$ Department of Physics and IPAP, Yonsei University, Seoul 120-479, Korea}
\maketitle
\begin{abstract}
We examine the charmless hadronic $B$ decay modes in the context
of $R$-parity violating ($\rpv$) supersymmetry.
We try to explain the large branching ratio (compared to
the Standard Model (SM) prediction) of the decay
$B^{\pm}\rightarrow \eta' K^{\pm}$. There exist data for other observed
$\eta^{(\prime)}$ modes and among these modes, the decay $B^{0}\rightarrow \eta
K^{*0}$  is also found to be large compared to the SM prediction. We investigate
all these modes and find that only two pairs of $\rpv$ coupling can satisfy the
requirements without affecting the  other
$B\ra PP$ and $B\ra VP$ decay modes barring the decay 
$B\ra\phi K$.  From this analysis, we determine the  preferred values of the
$\rpv$ couplings and the effective number of color $N_c$.
We also calculate the  CP asymmetry for the observed decay modes
affected by these new couplings.
\end{abstract}

\newpage
\section{Introduction}

For the  last few years, different exeperimental groups
have been accumulating plenty of  data for the charmless hadronic B decay modes.
The CLEO \cite{cleo,cleo2,cleo3,cleo4,cleo5}, the Belle
\cite{belle,belle2,belle3} and  the BaBar  collaboration
\cite{babar,babar2,babar3,babar4} are providing us with the information on the 
branching ratio (BR) and the CP asymmetry for different decay modes.   A clear
picture is about to emerge from these information.

Among the $B\ra PP$ ($P$ denotes a pseudoscalar meson) decay modes, the
branching ratio for  the decay $\bpetapk$  is found to be still larger than that
expected  within the Standard Model (SM). The SM contribution is about $3\sigma$
smaller than the experimental world average (see Fig.1).  
Among the $B\ra VP$ ($V$ denotes a vector meson) decay modes, 
the experimentally observed BR for the decay 
$B^{0}\rightarrow \eta K^{*0}$ has been aloso found to be $2\sigma$ 
larger than the SM. The decay $\bpphik$ has  been observed
recently, and   the BR for the newly observed decay
$B^{\pm}\rightarrow \eta K^{*\pm}$ is also now available.

In this paper, we address these large BR problems of 
$B^{\pm (0)}\rightarrow \eta^{(\prime)} K^{\pm (*0)}$ systems using
$R$-parity violating ($\rpv$) supersymmetric theories (SUSY).  The effects of
$\rpv$ couplings on $B$ decays have been investigated previously in the
literatures \cite{rparityb,dkc}, where  attempts were made to fit just the large BR for
$\bpetapk$\cite{dkc}. At present, we  have many more available results. 
Some of these results are concerned with decay modes involving
$\eta^{(')}$ and these modes are influenced by the same $\rpv$ 
coupling that affects 
$\bpetapk$. For example, the decay modes  
$B^{\pm}\rightarrow \eta K^{*\pm}$,  $B^{0}\rightarrow \eta K^{*0}$,  
$B^{0}\rightarrow \eta' K^{0}$  are affected by the new couplings which cure the
large BR problem of
$\bpetapk$. Hence,  it is natural to investigate  these newly observed  decay
modes and try to see whether all the available data can  be explained
simultaneously. We also
need to be concerned about the other observed (not involving
$\eta^{(\prime)}$) $B\ra PP$ and $B\ra VP$ decay modes, which could be
influenced by these new couplings. Our effort is not to affect the other modes
as much as possible, since except for  $B\ra \eta^{(\prime)}K^{(*)}$ decay
modes, the other observed modes fit the available data well\cite{fit,do}
within the SM.
Further, using the preferred values of different parameters (e.g., new couplings
etc.), we also make predictions for CP asymmetrey for these observed modes which
will be verified in the near future.   

We organize this letter as follows.  In section II, we give a very brief
introduction to the SM and $\rpv$ Hamiltonian, and list the  possible $\rpv$ 
operators  that can contribute to charmless $B$ decays. We  discuss the $B\ra PP$
and $B\ra PV$ decay modes  in section III. The new physics contributions to
different decay modes are also discussed. In section IV, we show  how $\rpv$ can
explain the branching ratio of $B\ra \eta^{(\prime)}K^{(*)}$ decay modes without
jeopardizing   many other $B\ra PP$ and $B\ra VP$ decay modes. 
 We also discuss the CP asymmetry of these decay modes. We conclude in section V.

\section{ Effective Hamiltonian for charmless decays } 

The effective Hamiltonian for charmless nonleptonic $B$ decays can be  written as
\be 
{\cal H} = {G_F\over\rtwo}\Big[ V_{ub}V^*_{uq}\sum_{i=1,2} c_iO_i
   - V_{tb}V^*_{tq}\sum_{i=3}^{12} c_iO_i\Big] + h.c.
\ee
The Wilson coefficients (WCs), $c_i$, contain the short-distance QCD
corrections. We find all our expressions in terms of the effective WCs and  
refer the reader to  the papers \cite{buras,ciuchini,desh,ali}   for a detailed 
discussion.  We use the effective WCs for the processes
$b\ra s\bar qq'$ and $b\ra d\bar qq'$ from Ref. \cite{desh}.  The regularization
scale is taken to be $\mu=m_b$.   In our  discussion, we will neglect small
effects  of the  electromagnetic moment operator
$O_{12}$, but will take into account  effects from the four-fermion operators
$O_1-O_{10}$ as well as the chromomagnetic operator $O_{11}$.

The $\rpv$ part of the superpotential of the minimal supersymmetric standard
model  (MSSM) can contain terms of the form
\be 
 {\cal W}_{\rpv}=\kappa_iL_iH_2 + \l_{ijk}L_iL_jE_k^c + \l'_{ijk}L_iQ_jD_k^c
          + \l''_{ijk}U_i^cD_j^cD_k^c \, 
      \label{superpot}
\ee   
where $E_i$, $U_i$ and $D_i$ are respectively the $i$-th type of  lepton,
up-quark and down-quark singlet superfields, $L_i$ and
$Q_i$ are the SU$(2)_L$ doublet lepton and quark superfields, and
$H_2$ is the Higgs doublet with the appropriate hypercharge.   From the symmetry
reason, we need $\l_{ijk}=-\l_{jik}$ and 
$\l''_{ijk}=-\l''_{ikj}$. The bilinear terms  can be rotated away with
redefinition of lepton and  Higgs superfields, but the effect reappears as
$\l$s, $\l'$s and  lepton-number violating soft terms \cite{roy}.  The first
three terms of Eq.(\ref{superpot}) violate the lepton number, whereas the fourth
term violates the baryon number.  We do not want all these terms to be  present
simultaneously due to catastrophic rates for proton decay.   In order to prevent
proton decay, one set needs to be forbidden.  

For our purpose, we will assume only  $\l'-$type  couplings to be present. 
Then, the
effective Hamiltonian for charmless nonleptonic
$B$ decay can be written  as
\be
\barr{rcl}
\dis \hlp (b\ra \bar d_j d_k d_n)
   & = & \dis d^R_{jkn} [ \bar d_{n\alpha} \gamma^\mu_L d_{j\beta}
                           \; \bar d_{k\beta} \gamma_{\mu R} b_{\alpha}] 
            + d^L_{jkn} [ \bar d_{n\alpha} \gamma^\mu_L b_{\beta}
                           \; \bar d_{k\beta} \gamma_{\mu R} d_{j\alpha}] 
              \ ,
             \\[1.5ex]
\dis \hlp (b\ra \bar u_j u_k d_n)
   & = & \dis u^R_{jkn} [\bar u_{k\alpha} \gamma^\mu_L u_{j\beta}
                           \; \bar d_{n\beta} \gamma_{\mu R} b_{\alpha}]
              \ ,
              \\[1,5ex]

\earr
    \label{rp_hamilt}
\ee    
with
\be
 \barr{rclcrclcl} d^R_{jkn} &=& \dis 
      \sum_{i=1}^3 {\l'_{ijk}\l'^{\ast}_{in3}\over 8\msnus},
      &  &  d^L_{jkn} &=& \dis \sum_{i=1}^3 {\l'_{i3k}\l'^{\ast}_{inj}\over
8\msnus},
      &  & (j,k,n=1,2)
           \\[1.5ex]  u^R_{jkn} &=& \dis \sum_{i=1}^3
{\l'_{ijn}\l'^{\ast}_{ik3}\over 8\msells},
      &  & 
      &  & 
      &  & (j,k=1, \ n=2),
\earr
\ee 
where $\alpha$ and $\beta$ are color indices and
$\gamma^\mu_{R, L} \equiv \gamma^\mu (1 \pm \gamma_5)$. The leading order QCD
correction to this operator is given by a scaling factor $f\simeq 2$ for
$m_{\tilde\nu}=200$ GeV.

The available data on low energy processes can be used to impose rather strict
constraints on many of these  couplings \cite{constraints,products,herbi}.  
Most such bounds have been  calculated under the assumption of 
there being  only one non-zero $\rpv$ coupling. 
There is no strong argument  to have only one $\rpv$  coupling
being nonzero. In fact,   a hierarchy of couplings may be  naturally
obtained \cite{products} on account of the mixings in either of the  quark and
squark sectors. In this paper, we  try to find out the values of  such $\rpv$
couplings for which  all available data are simultaneously satisfied. 
An important role will be
played by   the $\l'_{32i}$ -type couplings, the constraints on which are
relatively weak.

\section{$B\ra PP$ and $B\ra VP$ decay modes}

We consider next the  matrix elements of the various vector ($V_\mu$) and
axial vector ($A_\mu$) quark currents between  generic meson states.
For the decay constants of a pseudoscalar or a vector meson defined through
\be
\barr{rcl}
\bra 0|A_\mu|P(p)\ket & = & \dis if_P p_\mu \\
\langle 0|V_{\mu}|V(\epsilon,p)\rangle 
                      & = & f_{V} m_{V}\epsilon_{\mu} \ ,
\earr
\ee  
we use the followings (all values in MeV):
\be  
  f_\omega = 215,\ f_{K^*}  = 225,\ f_\rho = 215,\ f_\pi = 132,\ f_K  = 162,\
f_{\eta_1} = 146,\ f_{\eta_8} = 180.
\ee   
The decay constants of the mass eigenstates  $\eta$ and $\eta'$ are
related to those for the weak eigenstates through the relations
\[
\barr{rclcrcl} f^u_{\eta^{\prime}} & = &
\dis {f_8\over\sqrt{6}} \sin \theta
                     +{f_1\over\sqrt{3}} \cos \theta ,
     & \qquad & 
    f^s_{\eta^{\prime}} & = & \dis -2{f_8\over\sqrt{6}} \sin \theta
                              +{f_1\over\sqrt{3}} \cos \theta ,
      \\ f^u_{\eta} & = & \dis {f_8\over\sqrt{6}} \cos \theta 
                - {f_1\over\sqrt{3}} \sin \theta, 
     & \qquad & 
  f^s_{\eta} & = & \dis -2{f_8\over\sqrt{6}} \cos \theta
                            -{f_1\over\sqrt{3}} \sin \theta.
\earr
\] 
The mixing angle can be inferred from the  data on the
$\gamma\gamma$ decay modes\cite{angle} to be
$\theta \approx -22^\circ$.

The $B\ra P$  matrix element can be parameterized as
\be
\bra P(p')|V_\mu|B(p)\ket = \Big[ (p'+p)_\mu - {m_B^2-m_P^2\over q^2}q_\mu\Big ]
   F_1^{B\ra P} 
 + {m_B^2-m_P^2\over q^2}q_\mu F_0^{B\ra P} \ ,
\ee   
and the $B\ra V$ transition is given by
\be
\barr{l}
\langle V(\epsilon, p')|(V_{\mu}-A_{\mu})|B(p)\rangle
    \\
\dis  \hspace*{1em} =  
  {2 V \over {m_B+m_V}} 
     \epsilon_{\mu\nu\alpha\beta}\epsilon^{*\nu} p^{\alpha} p^{\prime\beta}
         \\ 
\dis  \hspace*{1em} +  
    i \left[ 
           (m_B+m_V) A_1 \: \epsilon^*_\mu + 
               \epsilon^* \cdot q 
                  \: \left\{- A_2 \frac{(p+p')_\mu}{m_B+m_V} 
                        + 2 m_V \frac{q_\mu}{q^2} \left( A_0 - A_3 
                                            \right)
                  \right\}
                  \right]~,
\earr
\ee 
with $$2m_VA_3 \equiv (m_B + m_V) A_1 - (m_B - m_V) A_2.$$   
The quantities $F_{0,1}^{B \ra P}$, 
$V$ and $A_{0,1,2,3}$ are the hadronic form factors and their values are given
in the next section.

The $\rpv$ part of the amplitude of  $B^{\pm}\ra\eta'K^{\pm}$ decay is 
\be
\barr{rcl}  {\cal M}_{\eta' K}^{\l'} &=& \dis
       \left( d^R_{121} - d^L_{112} \right) \xi A_{\eta'}^u 
        +  
          \left( d^L_{222} - d^R_{222} \right) \: 
                 \left[  \frac{\bar m}{m_s}
                   \left(A_{\eta'}^s -A_{\eta'}^u\right)- \xi  A_{\eta'}^s\right]
        \\[1.5ex]
       & + &  \dis 
        \left(d^L_{121} - d^R_{112} \right) \frac{\bar m}{m_d} A_{\eta'}^u
       + 
       u^R_{112} \left[\xi A_{\eta'}^u - {2 m_K^2 A_K \over (m_s+m_u) (m_b-m_u)} 
                 \right],
\earr
\ee 
where $\xi\equiv 1/N_c$ ($N_c$ denotes the effective number of color),
$\bar m \equiv m^2_{\eta'} / (m_b - m_s)$ and
\[ 
\barr{rcl} A_{M_1} & = & \bra M_2|J_b^\mu|B\ket \; \bra M_1|J_{l\mu}|0\ket ,\\ 
A_{M_2} & = &\bra M_1|J_b^{'\mu}|B\ket \; \bra M_2|J'_{l\mu}|0\ket.\earr
\]
$J$ and $J'$ stand for quark currents and the subscripts $b$ and 
$l$ indicate whether the current involves a $b$ quark or only the light quarks.
Analogous expressions hold for $B^{\pm}\ra\eta K^{\pm}$ where  we have to replace
$A^u_{\eta^{\prime}}$ by $A^u_{\eta}$, 
 $A^s_{\eta^{\prime}}$ by $A^s_{\eta}$ and $m_{\eta^{\prime}}$ by $m_{\eta}$.
Replacing a pseudoscalar meson by a vector meson, we
 also get similar expressions for the amplitudes of $B^{\pm(0)}\ra\eta'
K^{*\pm(0)}$ modes. The $\rpv$ part of the amplitude of 
$B\ra\phi K$ decay mode involves   only $d^L_{222}$ and  $d^R_{222}$, and 
\be
\barr{rcl}  {\cal M^{\l'}_{\phi K}} &=& \dis
         
          \left( d^L_{222} + d^R_{222} \right) \: 
                 \left[  \xi A_{\phi}\right],
\earr
\ee 
where $A_{\phi}=\bra K|J_b^\mu|B\ket \;\bra \phi|J_{l\mu}|0\ket$.

\section{Results} 

In our calculation, we use the following input for the Cabibbo-Kobayashi-Maskawa
(CKM) angles:
\begin{eqnarray}
|V_{cb}|&=&|V_{ts}|=0.04,~~ |V_{ub}|=0.087|V_{cb}|,~~
|V_{us}|=|V_{cd}|=0.218, \\ \nonumber |V_{ud}|&=&0.9722,~~ |V_{cs}|=0.9740,~~
|V_{tb}|=0.9988, \\ \nonumber
\beta&=&26^0,~~~\gamma=110^0 ~(80^0)~.
\end{eqnarray}
We first try to explain the large branching ratio of
$B^{\pm}\ra\eta'K^{\pm}$. The observed BR for this  mode in three different
experiments are \cite{cleo4,belle2,babar4}
\begin{equation}
{\mathcal B}( B^{\pm}\ra\eta'K^{\pm})\times
10^{6}=80^{+10}_{-9}\pm 7 \, ({\rm CLEO}),~~70\pm 8\pm 5\, ({\rm
BaBar}),~~79^{+12}_{-11}\pm 9 \, ({\rm Belle}). 
\end{equation}
The three results are close and we use the world average of them:
$(75 \pm 7) \times 10^{-6}$. 
The maximum BR in SM that we find is $42 \times 10^{-6}$
(Fig. 1). In the $\rpv$ SUSY framework, we find that the positive values of
$d^R_{222}$ and negative values of 
$d^L_{222}$ can increase the BR, keeping most of the other $B\ra PP$ and
$B\ra VP$ modes unaffected. The other $\rpv$ combinations are either  not enough
to increase in the BR or affect too many other modes. We divide our results into
two cases;\\
\noindent 
{\bf Case 1}: we use only $d^{R}_{222}$ (positive values) and \\
\noindent 
{\bf Case 2}: we  use a combination of 
$d^{R}_{222}$ (positive values) and $d^{L}_{222}$ (negative values). 

Let us start with {\bf Case 1}.   We first discuss the case of $\gamma=110^0$.
In this scenario we use $m_s~ ({\rm at}~ m_b ~{\rm scale}) =85$ MeV. 
In Fig. 1, we plot the BR for the decay 
$B^{\pm}\ra\eta^{\prime} K^{\pm}$  as a function of
$\xi$. We have used  $|\l'_{323}|=|\l'_{322}|=0.04$, 0.06, 0.08 and $m_{\rm
susy}=200$ GeV. We take $d^R_{222}$ to be positive.
The large branching ratio can be explained for
$\l'\geq0.05$. In our calculation, we use the following form factors:
\begin{equation}
A_0^{B\ra \omega}=A_0^{B\ra\rho}= A_0^{B\ra K^*}=0.45,\,\,\,
\,\,\, F^{B\ra K}_{0,1}=F^{B\ra \eta_1}_{0,1}=F^{B\ra
\eta_8}_{0,1}=0.29,\,\,\, F^{B\ra \pi^{\pm}}_{0,1}=0.26.
\end{equation}

\begin{figure}[htb]
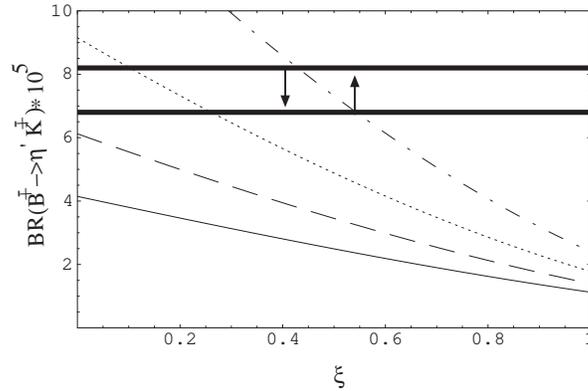

\centerline{ \DESepsf(fig1.epsf width 8 cm) }
\caption {\label{fig1} The BR for the decay  $B^{\pm}\rightarrow \eta'K^{\pm}$ vs
$\xi$. The solid line is for the SM. The dashed, dotted and dot-dashed  lines 
correspond to $|\l'_{323}|=|\l'_{322}|=0.04$, 0.06, 0.08, respectively. The bold
solid lines indicate the experimental world average bound.}
\end{figure}

\begin{figure}[htb]
\centerline{ \DESepsf(fig2.epsf width 8 cm) }
\caption {\label{fig2} The BR for the deacy $\bpphik$ vs $\xi$. The solid line is
for the SM. The dashed, dotted and dot-dashed  lines  correspond to
$|\l'_{323}|=|\l'_{322}|=0.04$, 0.06, 0.08, respectively. The bold
solid lines indicate the experimental world average bound.}
\end{figure}

\begin{figure}[htb]
\centerline{ \DESepsf(fig3.epsf width 8 cm) }
\caption {\label{fig3} The BR for the deacy $B^{\pm}\rightarrow \eta K^{*\pm}$
vs $\xi$. The solid line is for the SM. The dashed, dotted and dot-dashed 
lines  correspond to $|\l'_{323}|=|\l'_{322}|=0.04$, 0.06, 0.08, respectively.
The bold solid lines indicate the experimental bound.}
\end{figure}

\begin{figure}[htb]
\centerline{ \DESepsf(fig4.epsf width 8 cm) }
\caption {\label{fig4} The BR for the deacy $B^{0}\rightarrow \eta K^{*0}$ vs
$\xi$. The solid line is for the SM. The dashed, dotted and dot-dashed  lines 
correspond to $|\l'_{323}|=|\l'_{322}|=0.04$, 0.06, 0.08, respectively.
The bold solid lines indicate the experimental bound.}
\end{figure}

In Fig. 2, we plot the BR for  the deacy $B^{\pm}\ra \phi K^{\pm}$ for the same
set of couplings. The observed BR of this  mode by the CLEO \cite{cleo5} is
(in $10^{-6}$) $5.5^{+2.1}_{-1.8}\pm 0.6$.
The Belle and the BaBar collaboration have also observed
this mode \cite{belle3,babar3} with BRs (in $10^{-6}$)
$11.2^{+2.2}_{-2.0}\pm1.4$ and $7.7^{+1.6}_{-1.4}\pm0.8$, respectively.
{}From the figure we see that the BR is increasing with $|\l'|$.
The BR for $|\l'_{323}|=|\l'_{322}|=0.06$ and $\xi \simeq 0.25 - 0.3$ is
consistent with the world average bound $(7 - 9) \times 10^{-6}$.
Now combining Fig. 1 and Fig. 2, we find that
$|\l'_{323}|=|\l'_{322}|=0.06$ is allowed by both decay modes for
$\xi\sim 0.25$.
We note that the BaBar's number for this mode is quite close to the value
observed by the CLEO, but
the Belle's number is more than 2$\sigma$ away from the CLEO's central value.
We hope that this discrepancy will be sorted out in the near future.

In Fig. 3, we exhibit the BR for the deacy $B^{\pm}\rightarrow \eta K^{*\pm}$ as
a function of
$\xi$. The observed BR of this  mode \cite{cleo4} is
${\mathcal B}( B^{\pm}\ra\eta K^{*\pm})\times 10^{-6}=26.4^{+9.6}_{-8.2}\pm 3.3$.
We find that the solution we have got from our previous two decay modes holds in
this case: i.e., $|\l'_{323}|=|\l'_{322}|=0.06$ is allowed for $\xi\sim 0.15
-0.25$. 

Since all the parameters are fixed, now it is interesting to see whether the
decay $B^{0}\rightarrow \eta K^{*0}$ fits in the allowed region.
The observed BR for this  mode by CLEO collaboration \cite{cleo4} is
(in $10^{-6}$) $13.8^{+5.5}_{-4.6}\pm 1.6$.
The Belle and the BaBar collaboration have also
observed this mode \cite{belle3,babar2} with BRs (in $10^{-6}$)
$19.8^{+6.5}_{-5.6}\pm1.7$ and $21.2^{+5.4}_{-4.7}\pm 2.0$, respectively.
The SM BR is very small and cannot explain the experimental data.
In Fig. 4, we plot $B^{0}\rightarrow \eta K^{*0}$, where the world average value of
the data is expressed as the bold solid line.
We find that the dotted line ($|\l'|=0.06$) is allowed for $\xi \lesssim 0.2$.
But, the estimated BR for $|\l'_{323}|=|\l'_{322}|=0.06$ at
$\xi \simeq 0.25$ is just below and very close to the lower bound of the
average data.  In fact, $|\l'_{323}|=|\l'_{322}|=0.06$ is allowed by both the
CLEO and the Belle data for $\xi \lesssim 0.25$.  Only the BaBar's number is a
little bit larger than our estimated BR at $0.2 < \xi \lesssim 0.25$.

There exists a result for another decay  mode involving $\eta'$, i.e.,
$B^{0}\rightarrow \eta' K^{0}$. For $|\l'|=0.06$  
and $\xi=0.25$ we find the BR is $88.3\times 10^{-6}$. 
The CLEO bound is (in $10^{-6}$) $89^{+18}_{-16}\pm 9$. The Belle and
the BaBar collaboration also have reported observation of this mode but with
less significance, and their results are (in $10^{-6}$)
$55^{+19}_{-16}\pm 8$  and $42^{+13}_{-12}\pm 4$, respectively. We see that the
results for this mode from the three experiments are not consistent. Our result
in this scenario can be taken as a prediction which is in agreement with the
CLEO result.

\begin{table}
\caption{The branching ratios $({\mathcal B})$  for  $B$ decays into $PP$ and
$VP$ final states at $\xi =0.25$.}
\begin{tabular}{|c|c|c|c|}
Decay modes&${\mathcal B}$ $(10^{-6})$&
${\mathcal B}$ $(10^{-6})$&Experimental ${\mathcal B}$
$(10^{-6})$ \\
&$\gamma=110^0$&$\gamma=80^0$&\\\hline
$B^0 \rightarrow \pi^+ \pi^-$&5.96&6.0&$4.3^{+1.6}_{-1.4}\pm 0.5$ {(\rm
CLEO\cite{cleo})}, $5^{+2.3+0.4}_{-2.0-0.5}$ {(\rm
Belle\cite{belle})}\\&&&$4.1\pm 1\pm 0.7$ {(\rm BaBar\cite{babar})}\\\hline
$B^+ \rightarrow  K^+\pi^0$&9.59&9.19&$11.6^{+3.0+1.4}_{-2.7-1.3}$\cite{cleo},
 $16.3^{+3.5+1.6}_{-3.3-1.8}$ \cite{belle}\\
 &&&$10.8^{+2.1}_{-1.9}\pm 1.0$\cite{babar}\\\hline
$B^+ \rightarrow K^0 \pi^+$&13.6&15.63&$18.2^{+4.6}_{-4.0}\pm 1.6$\cite{cleo},
 $13.7^{+5.7+1.9}_{-4.8-1.8}$ \cite{belle}\\
 &&&$18.2^{+3.3}_{-3.0}\pm 2.0$\cite{babar}\\\hline
$B^0 \rightarrow K^+ \pi^-$&15.10&14.50&$17.2^{+2.5}_{-2.4}\pm 1.2$\cite{cleo},
 $19.3^{+3.4+1.5}_{-3.2-0.6}$ \cite{belle}\\
 &&&$16.7\pm 1.6\pm 1.3$\cite{babar}\\\hline
$B^0 \rightarrow K^0 \pi^0$&8.68&9.87& $14.6^{+5.9+2.4}_{-5.1-3.3}$\cite{cleo},
 $16.0^{+7.2+2.5}_{-5.9-2.7}$ \cite{belle}\\
 &&&$8.2^{+ 3.1}_{-2.7}\pm 1.2$\cite{babar}\\\hline\hline
$B^+ \rightarrow \omega \pi^+$&9.518&9.0&$11.3^{+3.3}_{-2.9}\pm 1.5$ {(\rm
CLEO\cite{cleo2})}, $6.6^{+ 2.1}_{-1.8}\pm 0.7${(\rm BaBar\cite{babar2})}\\\hline
$B^+ \rightarrow \rho^0 \pi^+$&9.418&9.45&$10.4^{+3.3}_{-3.4}\pm 2.1${\cite{cleo2}},
$24\pm 8\pm 3${\cite{babar2}}\\\hline
$B^0 \rightarrow \rho^\mp \pi^\pm$ (sum)&35.23&35.03&$27.6^{+8.4}_{-7.4}\pm
4.2${\cite{cleo2}}\\\hline
$B^+ \rightarrow \omega h$&10.87&12.18&$14.3^{+3.6+2.1}_{-3.2-2.1}$ {(\rm
CLEO\cite{cleo3})}\\
$B^0 \rightarrow K^{*+} \pi^-$& 5.93&4.15& $22^{+8+4}_{-6-5}${\cite{cleo3}}
\end{tabular}
\end{table}

The ${\cal B}(B\rightarrow X_s \nu\nu)$ can put bound on
$\l'_{322}\l'^{\ast}_{323}$ in certain limits. Using the Refs. \cite{grossman} and
the  experimental limit (BR$<6.4\times 10^{-4}$) on the above BR \cite{ALEPH}, we
find that $\l'\leq 0.07 $. However, if we go to any realistic scenario, for
example grand unified models (with $R$ parity violation), we find a natural
hierarchy among the sneutrino and squark masses. The squark masses are much
heavier than the sneutrino masses and the bound does not apply any more.

The other observed $B\ra PP$ and $B\ra VP$ decay modes are listed in Table I for
$\xi=0.25$ and we find that the BRs are within the experimental limits. Our
result on $B^0 \rightarrow K^{*+} \pi^-$ is compatible within 2$\sigma$ range,
but this measurement still involves large error.

We can use smaller value of $\gamma$, e.g., $\gamma=80^0$ to fit the
$B\ra \eta^{(\prime)} K^{(*)}$ data. In this scenario we use
$m_s~ ({\rm at}~ m_b ~{\rm scale}) =75$ MeV. In Table II we show the BRs for $B\ra
\eta^{(\prime)} K^{(*)}$ and  in Table I we show the BRs for the other observed
$B\ra PP$ and $B\ra VP$ decay modes in the case of $\gamma=80^0$. Again, we find
the fit is reasonable. In this case, we use $A_0^{B\ra K^*}=0.4$ and keep the
other inputs unchanged.

\begin{table}
\caption{The branching ratios $({\cal B})$ and the CP asymmetries for $B
\ra\eta^{(\prime)} K^{(*)}$ and $B \ra \phi K$.}
\begin{tabular}{|c|cc|cc||cc|cc|}
 & $\delta =0$, &
 & $\delta = 15^0$&  &$\delta = 0$, &
 & $\delta = 55^0$&\\
 & $\gamma =110^0$, &
 & $\gamma =110^0$&& $\gamma =80^0$, &
 & $\gamma =80^0$&\\ mode & ${\cal B} \times 10^{6}$ & ${\cal A}_{CP}$

& ${\cal B} \times 10^{6}$ & ${\cal A}_{CP}$& ${\cal B} \times 10^{6}$ & ${\cal
A}_{CP}$

& ${\cal B} \times 10^{6}$ & ${\cal A}_{CP}$
\\ \hline

$B^+ \to \eta^{\prime} K^+$ & 68.9 & 0.01 & 68.3 & 0.04 & 82.1 & 0.01 & 68.3 &
0.11
\\ $B^+  \to \eta K^{*+}$ &  36.4 & 0.03 & 36.4 & 0.04 & 36.5 & 0.03 & 32.7 &
0.09
\\ $B^0  \to \eta' K^0$ & 88.3 & 0.00 & 86.8 & 0.03 & 110.2 & 0.00 & 87.1 & 0.12
\\ $B^0  \to \eta K^{*0}$ & 14.0 & $-0.39$ & 14.6 & $-0.42$& 14.8 & $-0.28$ &
20.4 & $-0.56$
\\ $B^+  \to \phi  K^+$ & 7.11 & 0.00 & 6.97 & 0.04& 7.10 & 0.00 & 5.76 & 0.14
\end{tabular}
\end{table}

We now calculate the CP asymmetry for different $B \ra\eta^{(\prime)} K^{(*)}$
modes. The CP asymmetry, ${\mathcal A}_{CP}$, is defined by
\begin{eqnarray} 
{\mathcal A}_{CP} = {{\mathcal B}(B \rightarrow f)  -{\mathcal
B}(\overline B \rightarrow \bar f) \over {\mathcal B}(B \rightarrow f) +{\mathcal
B}(\overline B \rightarrow \bar f)}~,
\end{eqnarray} 
where $B$ and $f$ denote a $B$ meson and a generic final state,
respectively. Let us define
\begin{equation}
\l'_{323} \l'^{\ast}_{322} =|\l'_{323}
\l'^{\ast}_{322}|e^{i\delta},
\end{equation}
where $\delta$ denotes the phase difference between $\lambda_{323}^{\prime}$
and $\lambda_{322}^{\prime}$.
So far we have discussed the $\delta=0^0$ situation.
In Table II, we calculate the CP asymmetries for $B\ra\eta K$ 
modes  for different values of
$\delta$ and $\gamma$.  The maximum values of $\delta$ allowed by the BR of
$B^{\pm} \ra\eta' K^{\pm}$ are $\delta=15^0$ for $\gamma=110^0$ and
$\delta=55^0$ for $\gamma=80^0$. We find that ${\cal A}_{CP}$ is very large for
$B^0  \to \eta K^{*0}$ mode and is predicted to be $-39\%~(-42\%)$ for
$\delta=0^0~(15^0)$,
$\gamma=110^0$, and $-28\%~(-56\%)$ for $\delta=0^0~(55^0)$,
$\gamma=80^0$. ${\cal A}_{CP}$ for $B^{\pm} \ra\phi K^{\pm}$ is large ($14\%$)
for  $\delta=55^0$ and $\gamma=80^0$. The other $\eta^{(\prime)}$ modes are also
found to be large ($\sim 10\%$) for the above set of parameters.

\noindent{\bf Case 2}: We use the same form factors as in the case $\gamma = 80^0$
of {\bf Case 1} and we use
$m_s~ ({\rm at}~ m_b ~{\rm scale}) =85$ MeV and $\gamma=110^0$. We now use the
combination of $d^R_{222}$ and $d^L_{222}$. We assume $d^R_{222}=-d^L_{222}$.
In this scenario, the $\rpv$ coupling part of the amplitude in $B\ra\phi K$
decay mode canceled exactly (Eq. 10). (In fact, our solution still works when
the cancellation is incomplete by about 5$\%$.)
But we still have contributions to $B^{\pm}\ra\eta' K^{\pm}$ (Eq. 9)
and to increase the BR we choose
$d^R_{222}$ to be positive. There is no $\rpv$ contribution to the other $B\ra
PP$ and $B\ra VP$ modes in this case as well.

\begin{figure}[htb]
\centerline{ \DESepsf(fig5.epsf width 8 cm) }
\caption {\label{fig5} The BR for the deacy $B^{\pm}\rightarrow \eta'K^{\pm}$ vs
$\xi$. The solid line is for the SM. The dashed, dotted and dot-dashed  lines 
correspond to $|\l'|=0.035$, 0.052, 0.07, respectively. The bold solid lines
indicate the experimental bound.}\end{figure}

\begin{figure}[htb]
\centerline{ \DESepsf(fig6.epsf width 8 cm) }
\caption {\label{fig6} The BR for the deacy $B^{\pm}\rightarrow \eta K^{*\pm}$
vs $\xi$. The solid line is for the SM. The dashed, dotted and dot-dashed 
lines  correspond to $|\l'|=0.035$, 0.052, 0.07, respectively. The bold solid
lines indicate the experimental bound.}\end{figure}

\begin{figure}[htb]
\centerline{ \DESepsf(fig7.epsf width 8 cm) }
\caption {\label{fig7} The BR for the deacy $B^{0}\rightarrow \eta K^{*0}$ vs
$\xi$. The solid line is for the SM. The dashed, dotted and dot-dashed  lines 
correspond to $|\l'|=0.035$, 0.052, 0.07, respectively. The bold solid lines
indicate the experimental bound.}\end{figure}

In Fig. 5, we plot the BR for the deacy
$B^{\pm}\ra\eta' K^{\pm}$  as a function of
$\xi$. We have used  $|\l'|=0.035$, 0.052, 0.07, and $m_{\rm susy}=200$ GeV. In
this case, the large branching ratio can be explained for
$\l'\geq0.05$.

In Fig. 6 and Fig. 7, we plot the BRs for $B^{\pm}\ra\eta K^{*\pm}$ and
$B^{0}\ra\eta K^{*0}$.
Combining Figs. 5, 6 and 7, we find that $|\l'|$=0.052 and $\xi\simeq0.4-0.6$ can
explain all the data.
In this scenario our result on $B^0 \to \eta K^{*0}$ is allowed by the world
average bound.
The BR of  $B^{0}\rightarrow \eta' K^{0}$  
is $107.4\times 10^{-6}$ for $|\l'|=0.052$  and $\xi=0.45$ and is
allowed by the CLEO data.

The BR for the deacy $B^{\pm}\ra\phi K^{\pm}$ does not have a $\rpv$
contribution due to the cancellation. The SM line (solid) in Fig. 2 needs to be
used in this case and we find that $\xi\simeq0.45$ is allowed.  The BRs of the
other observed
$B\ra PP$ and $B\ra VP$ modes do not get affected by the new couplings and these
modes seem to fit the data reasonably well for  
$\xi\simeq 0.3-0.5$ \cite{do}.
We also calculate the ${\cal A}_{CP}$ for this case. Since we have assumed that
$d^R_{222}=-d^L_{222}$, the phase difference between $\lambda^{\prime}_{322}$ and
$\lambda^{\prime}_{332}$ is ($\delta+\pi$), where $\delta$ is the phase
difference between $\lambda^{\prime}_{323}$  and $\lambda^{\prime}_{322}$.

\begin{table}
\caption{The branching ratios $({\cal B})$ and the CP asymmetries for $B
\ra\eta^{(\prime)} K^{(*)}$ and $B \ra \phi K$.}
\begin{tabular}{|c|cc|cc|}
 & $\delta =0$ & & $\delta =20^0$ &
\\ mode & ${\cal B} \times 10^{6}$ & ${\cal A}_{CP}$ & ${\cal B} \times 10^{6}$
& ${\cal A}_{CP}$
\\ \hline
$B^+ \to \eta^{\prime} K^+$ & 69.3 & 0.01 & 68.0 & 0.05
\\ $B^+  \to \eta K^{*+}$ &  27.9 & 0.04 & 27.8 & 0.05
\\ $B^0  \to \eta' K^0$ & 107.4 & 0.00 & 104.5 & 0.05
\\ $B^0  \to \eta K^{*0}$ & 20.5 & $-0.71$ & 21.1 & $-0.72$
\\ $B^+  \to \phi  K^+$ & 6.56 & 0.00 & 6.56 & 0.00
\end{tabular}
\end{table}


In Table III, we calculate the BRs and the CP asymmetries for $B\ra
\eta^{(\prime)} K^{(*)}$ and $B \ra \phi K$ for different values of $\delta$.
The maximum value of $\delta$ allowed by the
BR  for 
$B^{\pm} \ra\eta' K^{\pm}$ is
$\delta=20^0$ for $\gamma=110^0$.  As
in $\bf Case 1$, we find that ${\cal A}_{CP}$ is very large for 
$B^0  \to \eta K^{*0}$ mode and is predicted to be $-71\%~(-72\%)$ for
$\delta=0^0~(20^0)$. The other $\eta^{(\prime)}$ modes are found to be $\sim
5\%$ for the above set of parameters.

\section{Conclusion}

We have studied $B \ra\eta^{(\prime)} K^{(*)}$ modes in the
context of $\rpv$  supersymmetric theories. We have isolated the necessary
$\rpv$ couplings ($\lambda^{\prime}_{323}
\lambda^{\prime *}_{322}$ and $\lambda^{\prime}_{322} \lambda^{\prime *}_{332}$)
that satisfy the experimental results for the BRs of these modes  reasonably
well. The Standard Model contribution is less than $2\sigma$ for some of these
modes. These new couplings do not affect any  other $B\ra PP$ and
$B\ra VP$ modes except for the decay  $B^{\pm}\ra\phi K^{\pm}$. We have
shown that the calculated BR for  $B^{\pm}\ra\phi K^{\pm}$ agrees with  the
experimental data.

We  found solutions for both large and small values of $\gamma=110^0,\,80^0$ and
two different values of $\xi(=1/N_c) \simeq 0.25,\,0.45$ 
for two different scenarios. For
our solutions, we need $|\l'| \sim 0.05-0.06$ for $m_{\rm susy}=200$ GeV. Using
these preferred values of the parameters, we   calculated the CP asymmetry of
different observed modes affected by the new couplings  and  found that the CP
asymmetry of $B^0  \to \eta K^{*0}$ is large  ($\sim 28\%-72\%$) and the CP
asymmetry of other modes also can be around $10-15\%$.

These new coupling can be also examined in the RUN II through the associated
production of the lightest chargino ($\chi^{\pm}_1$) and the second lightest
neutralino ($\chi^0_2$).
$\chi^{\pm}_1$ and $\chi^0_2$ decay into lepton plus 2 jets (for example), where
one of the jets is a $b$ jet. The final state of this production process
contains 2 leptons plus 4 jets. So the signal is quite interesting and unique. \\

\begin{center}
{\bf ACKNOWLEDGMENTS}
\end{center}

The work of C.S.K. was supported
in part by  CHEP-SRC Program, Grant No. 20015-111-02-2
and Grant No. R03-2001-00010 of the KOSEF,
in part by BK21 Program and Grant No. 2001-042-D00022 of the KRF,
and in part by Yonsei Research Fund, Project No. 2001-1-0057.
The work of B.D. was supported in part by National Science Foundation grant No.
PHY-0070964.
The work of S.O. was supported by Grant No. 2001-042-D00022
of the KRF.


\end{document}